\documentclass[a4paper]{jpconf}
\usepackage{graphicx}
\begin{document}
\title{Helioseismology with PICARD}

\author{T~Corbard$^1$, D~Salabert$^1$, P~Boumier$^2$, T~Appourchaux$^2$, A~Hauchecorne$^3$, P~Journoud$^2$, A~Nunge$^2$,  B~Gelly$^4$,
 J~F~Hochedez$^3$,  A~Irbah$^3$, M~Meftah$^3$, C~Renaud$^1$, S~Turck-Chi\`eze$^5$ }

\address{$^{1}$ Laboratoire Lagrange, UMR7293, 
 Universit\'e de Nice Sophia-Antipolis, CNRS, Observatoire de la C\^ote d'Azur,
 Bd. de l'Observatoire, 06304 Nice, France}

\address{$^{2}$ Institut d'Astrophysique Spatiale, CNRS-Universit\'e Paris XI, UMR 8617, 91405 Orsay Cedex, France}

\address{$^{3}$ Laboratoire Atmosph\`eres, Milieux, Observations Spatiales,CNRS,
 Universit\'e Paris~VI \& Universit\'e de Versailles Saint-Quentin-en-Yvelines,
 IPSL, F-78280 Guyancourt, France}

\address{$^{4}$ Themis, UPS 853 du CNRS, c/o IAC, Via Lactae s/n, 38200 La Laguna, Tenerife, Spain}

\address{$^{5}$ CEA/IRFU/Service d'Astrophysique, AIM, CE Saclay, 91191 Gif sur Yvette, France}

\ead{Thierry.Corbard@oca.eu}

\begin{abstract}
PICARD is a CNES micro-satellite launched in June 2010 \cite{Thuillier06}. Its main goal is to measure the solar shape, total and spectral irradiance during the ascending phase of the activity cycle. The SODISM telescope onboard PICARD  also allows us to conduct a program for helioseismology in intensity at 535.7 nm \cite{Corbard08}. One-minute cadence low-resolution full images are available for a so-called medium-$l$ program, and high-resolution images of the limb recorded every 2 minutes are used to study mode amplification near the limb in the perspective of g-mode search. First analyses and results from these two programs  are presented here.
\end{abstract}

\section{Medium-$l$ program in intensity}
Full continuum images at 535.7~nm (bandwidth 0.5~nm) are recorded by SODISM \cite{Meftah2013} every minute with a spatial resolution of  about 1 square arcsecond on a $2048^2$  CCD. For telemetry reasons, these images are downgraded to $256^2$  pixels (about 8x8 arcsecond) by a simple onboard binning before being transmitted to ground. We have analyzed a dataset covering 209 consecutive days (2011/04/16-2011/11/10 ) with a duty cycle of  $74.4\%$. The gaps are mostly due to interruptions for the other PICARD program (astrometric measurements at different wavelengths) and to the crossing of the South Atlantic Anomaly, which is the location of transient signals generated by trapped protons exposing low-altitude, Sun-synchronous orbiting satellites, like PICARD, to strong radiations.

\begin{figure}[t]
\begin{minipage}{14pc}
\includegraphics[width=14pc]{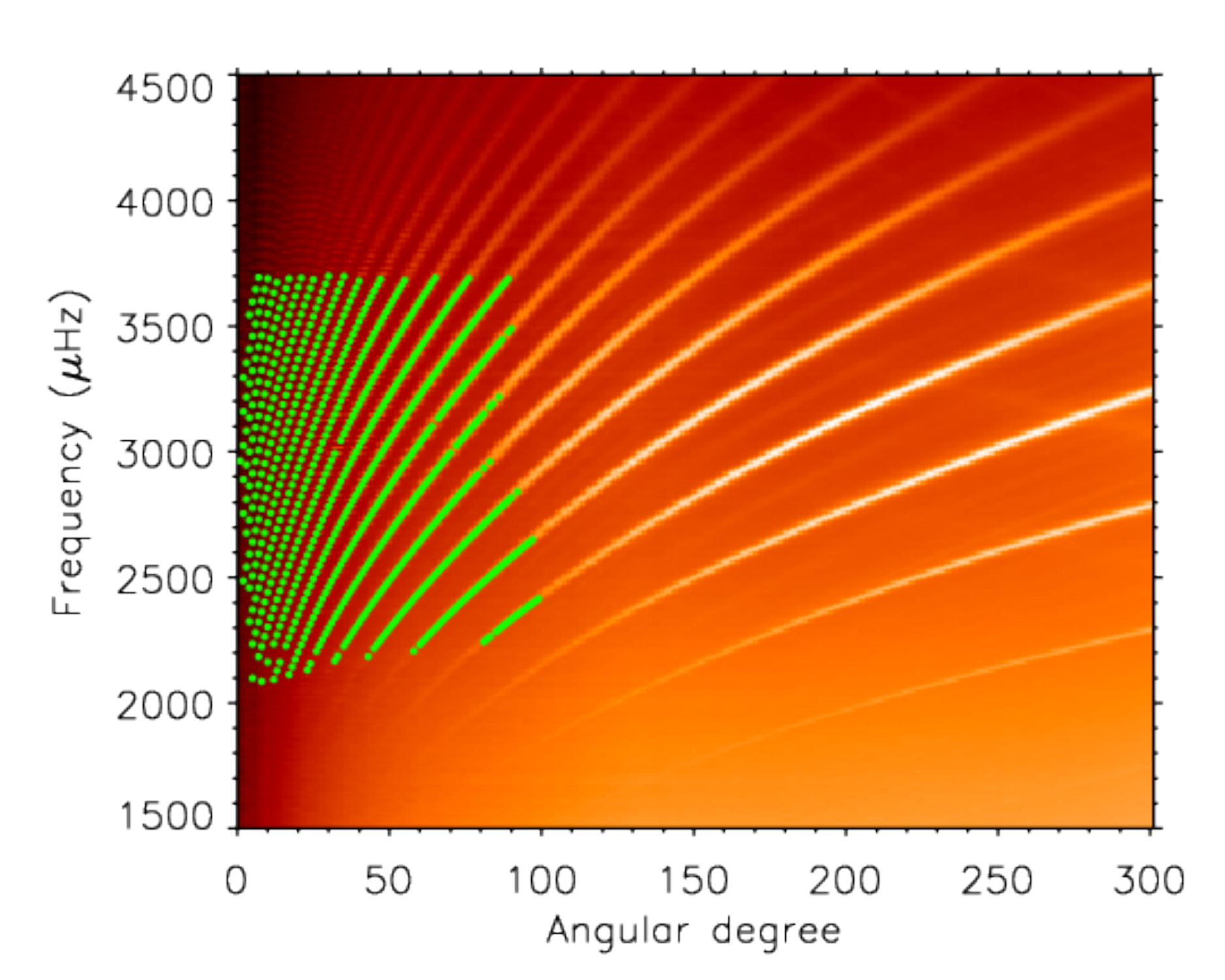}
\includegraphics[width=14pc]{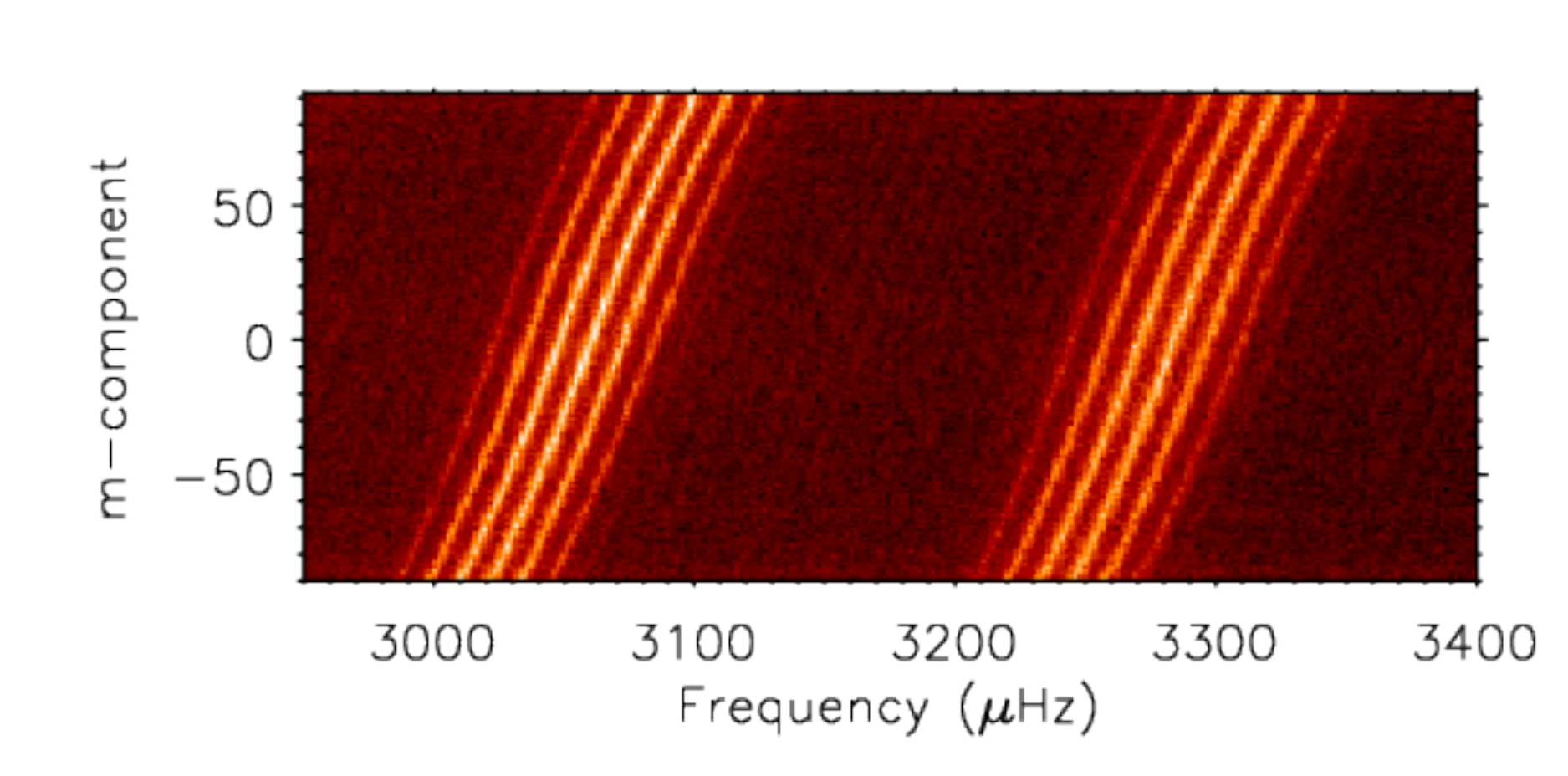}
\caption{\label{Fig:diags}Top: $l$-$\nu$ diagram  from 209-day PICARD observations. The fitted modes are shown by the green dots.
 Bottom:  $m$-$\nu$ diagram for $l$=90 and radial orders $n$=7 (left) and  $n$=8 (right). The $l$-leaks from $l$=90$\pm$3 are clearly seen.}
\end{minipage}\hspace{2pc}%
\begin{minipage}{22pc}
\includegraphics[width=22pc]{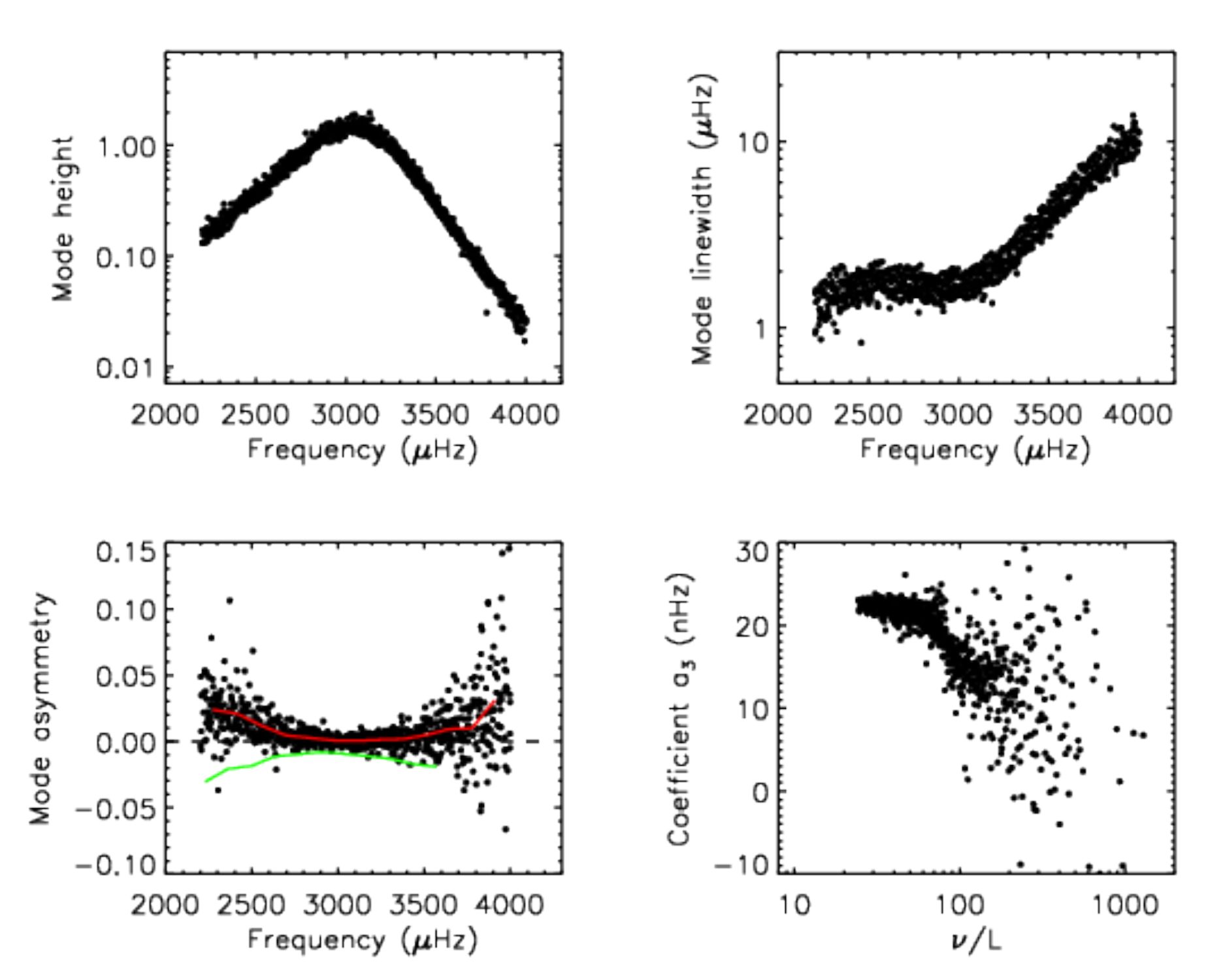}
\caption{\label{Fig:params}Fitted mode parameters: amplitude (upper left), linewidth (upper right), peak asymmetry (lower left), and $a_3$ splitting coefficients (lower right). The red line is a fit of SODISM positive asymmetries in intensity, while the green line shows the negative asymmetries obtained from the SOHO/GOLF velocity signal.}
\end{minipage} 
\end{figure}

Figure~\ref{Fig:diags} (top) shows the $l$-$\nu$ diagram  obtained  from this 209-day dataset for angular degree up to $l$=300. As expected from the instrumental resolution,  a spatial aliasing for spherical harmonics above $l$=200 is present. We can also see a temporal aliasing for frequencies above 4.5 mHz. This is the signature of a spurious cycle in the 2-min sampling slots corresponding to the astrometric program sequences. 
Figure~\ref{Fig:diags} (bottom)  illustrates a $m$-$\nu$ diagram for $l$=90 clearly showing the S-shape signature of the differential rotation. 
Figure~\ref{Fig:params} shows the  mode parameters modelled using the asymmetrical  line profile formula of Nigam and Kosovichev \cite{Nigam98b} and fitted using a Maximum Likelihood Estimator minimization  \cite{App98a, Chanothesis}. For a given mode ($n$,$l$), the 2$l$+1 $m$-components of the multiplets are fitted simultaneously  using  an expansion on orthogonal polynomials \cite{Schouthesis} with up to 9 so-called a-coefficients. The $l$ and $m$ mode leakages are included in the fitting process. Positive peak asymmetries are obtained as expected from previous analyses of intensity signal. The $a_3$ splitting coefficients plotted as a function of the mode turning point proxy $\nu/L$, where $L=\sqrt{l(l+1)}$, exhibit  a gradient which is a typical signature of the solar tachocline.  Improvements of the peak-fitting process are currently underway. We need for instance to better model all the components of the leakage matrix in order to fit higher angular degrees. 

\begin{figure}[!h]
\includegraphics[width=21pc]{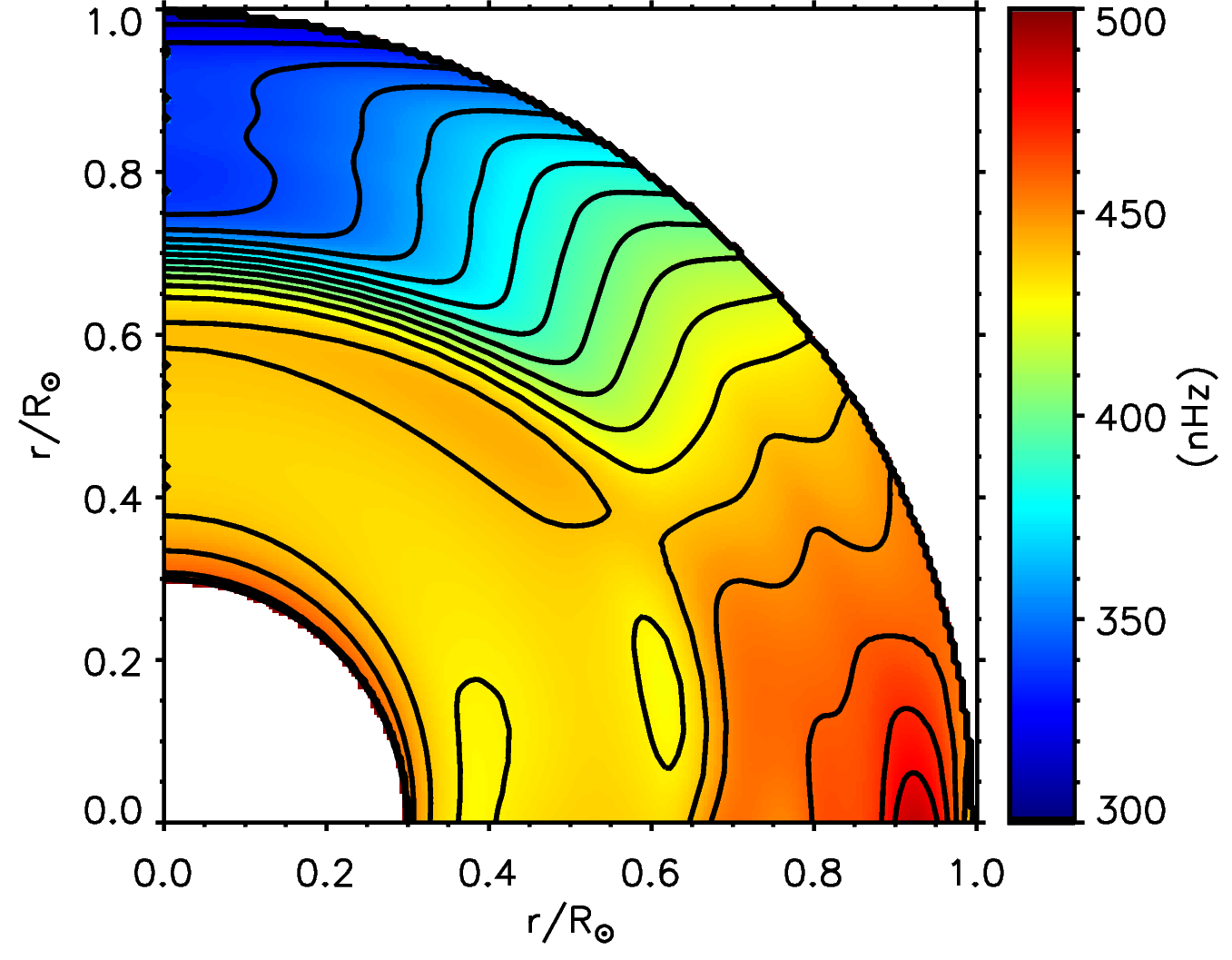}\hspace{1pc}
\begin{minipage}[b]{14pc}\caption{\label{Fig:rot}Symmetric part of the internal solar rotation rate with respect to the equator.  Iso-contours are shown every 10~nHz from 300~nHz up to 480~nHz. This was obtained by inverting the frequency splittings estimated  from PICARD/SODISM intensity images over the period from April 16 to  November 10 2011.
 The solution below 0.3$R_{\odot}$ is poorly constrained by the data and is not shown.\vspace{2pc}}
\end{minipage}
\end{figure}

Figure~\ref{Fig:rot} shows a Regularized Least-Square (RLS) inversion \cite{Corbard97} of the measured rotational splittings of 769 modes from $l$=1 to $l$=99. 
The steep gradient  of the tachocline at the base of the convection zone around 0.7~$R_{\odot}$ is clearly seen. The inferred rotation rate is similar to the one previously obtained from velocity data \cite{Corbard97, Rachel}: 
a radial gradient is
obtained close to the surface followed by a latitudinal differential rotation 
 persisting through the whole convection zone. The radiative 
interior is roughly rigid at a rate of $435\pm5$~nHz which corresponds 
to the surface rate for latitudes between $30$ and $40$ degrees. Finally, we note that a local maxima is found close to the equator 
near $0.95R_{\odot}$.  
These preliminary results are very encouraging. We are  
currently  working in order 
to extend the peak-fitting analysis towards higher $l$ 
and to increase the number of fitted $a$-coefficients thus improving
 the resolution of our inversion.
The  reversal of peak asymmetry observed between velocity and intensity 
power spectra is linked
to the effect of a background component correlated to the
modes \cite{Nigam98a}. The complete two years  of PICARD observations 
available to date should then be used to confront the details 
of internal rotation studies obtained
 using  simultaneous velocity and intensity data from SDO/HMI or GONG.
By fitting intensity, velocity,
their phase difference and coherence signals simultaneously,
we  hope to nail down systematics in either observables and 
better understand the correlated
background noise  \cite{Severino01, Barban04}. 
This will help  to better interpret the fitted frequencies
and frequency splittings in terms of internal
structure and dynamics.

\section{Limb seismology} 
Every 2 minutes, the full resolution of PICARD/SODISM images are kept in an annulus of 22 pixels around the solar limb (increased to 31 pixels since April 2012).  In April 2011 and April 2012, 3 consecutive days were dedicated to  helioseismic  measurements at 535.7~nm without interruption for the other PICARD objectives. These datasets with a duty cycle above 95$\%$ have been used in order to study the expected  mode amplification  at the extreme limb \cite{App98b, Toutain99, Toner99}.  Intensity data from SOHO/MDI and SDO/HMI were also used for comparison.

\begin{figure}[!h]
	\centerline{\includegraphics[width=0.45\textwidth]{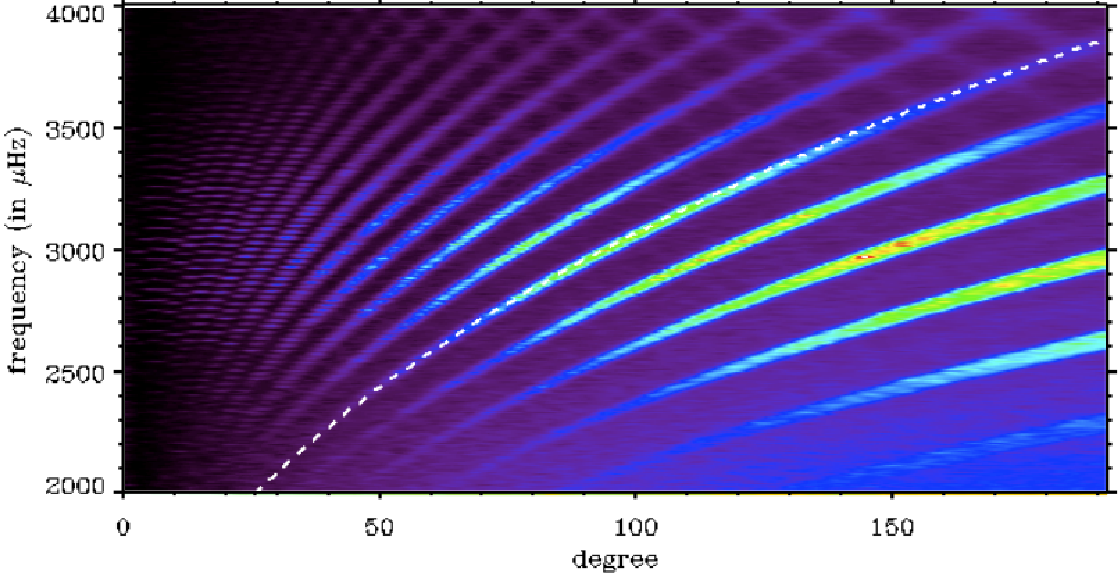}\hspace{1pc}\includegraphics[width=0.45\textwidth]{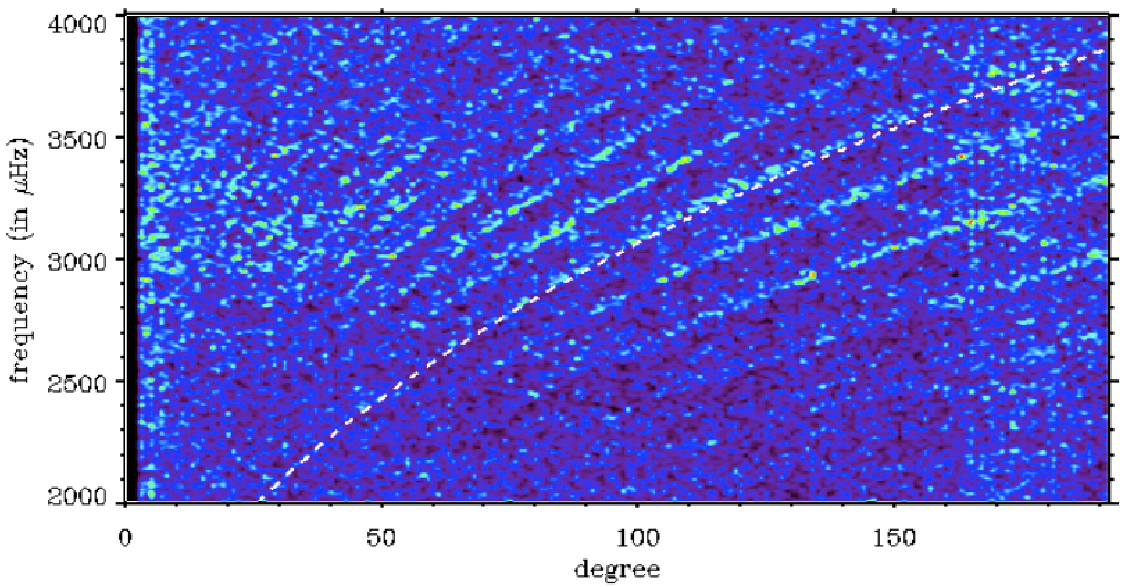}
}
	\caption{\label{Fig:complnu}Classical spherical harmonic transform of full images ($l$-$\nu$ diagram ) (left) and limb transform of limb pixels ($l^\prime$-$\nu$ diagram) (right). The theoretical $n$=6 ridge is indicated by a dashed line for both cases 
clearly showing a frequency shift between the two diagrams.}
\end{figure}

\begin{figure}
\centerline{\includegraphics[width=0.8\textwidth]{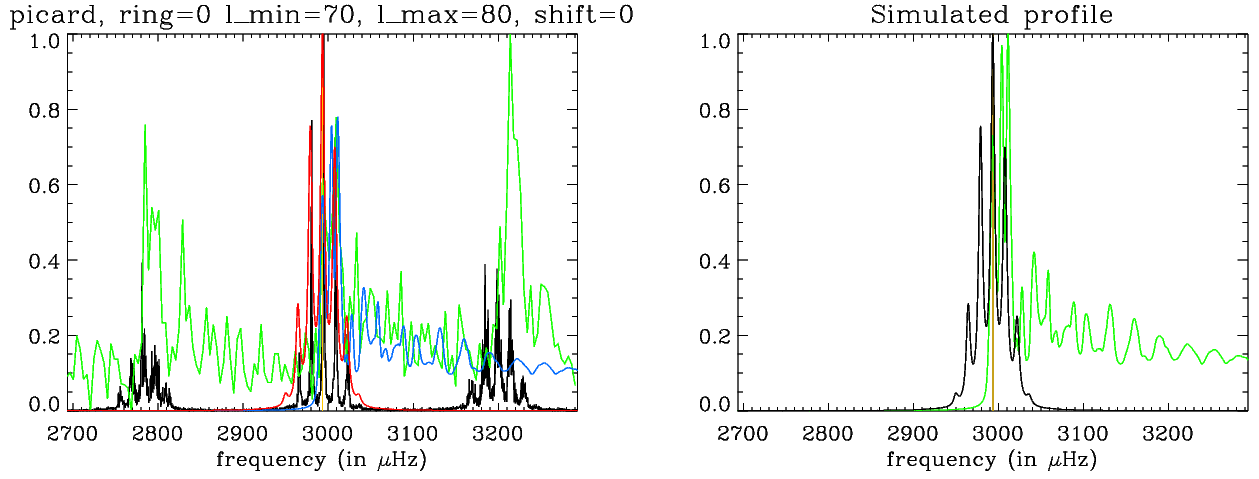}}
\centerline{\includegraphics[width=0.8\textwidth]{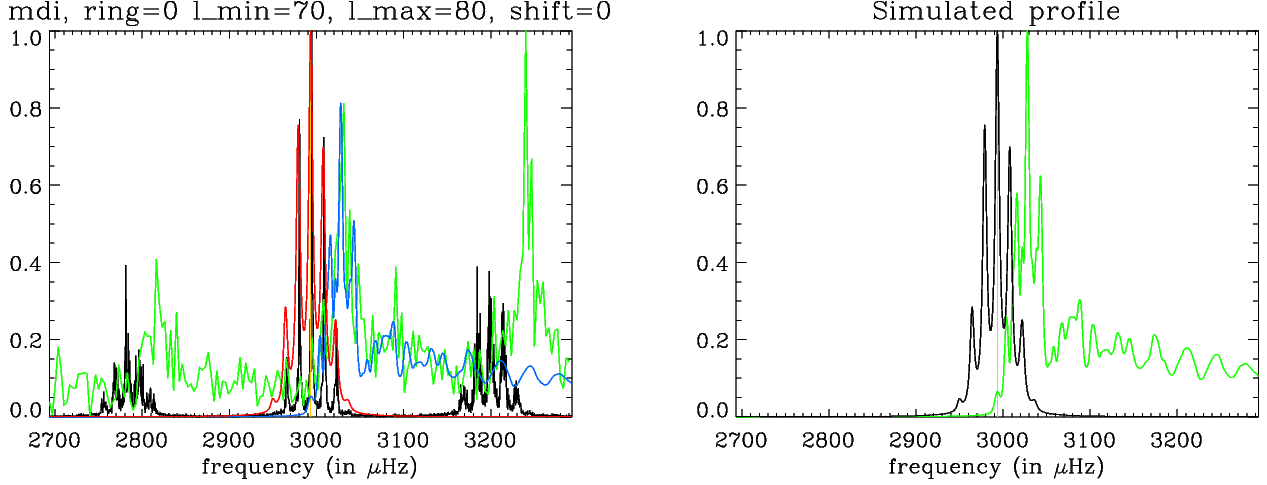}}
\caption{\label{Fig:simu}Comparison between observed  and simulated profiles. On the left panels, the black lines are ridges in SOHO/MDI $l$-$\nu$ diagrams while the green lines are ridges in the PICARD/SODISM (top) and SOHO/MDI (bottom) $l^\prime$-$\nu$ diagrams.  The two panels on the right show the simulated profiles for both  the
$l$-$\nu$ (black) and $l^\prime$-$\nu$ (green) diagrams which are also shown with an appropriate scalling
on the left panels as the red and blue lines respectively.}
\end{figure}

In order to analyse the limb signal, we calculated
the Fourier transform  $e^{il^\prime\rho}$ along the limb using the polar coordinates, where $\rho$=1 at the disk center and $\rho$=0 at the limb.
Figure~\ref{Fig:complnu} shows the comparison between the $l$-$\nu$ diagram obtained from a classical spherical harmonic transform of full images (left) and the $l^\prime$-$\nu$ diagram (right) resulting from the limb transform 
as described above. A frequency shift between ridges of the same radial order is clearly seen. 
We carried out simulations in order to understand this shift and the link between the two diagrams \cite{App02}. For a given $l^\prime$,
 it can be shown that the limb response are non zero for value of $l$ higher than 
$l^\prime$. Since the mode frequencies increase with the degree, the leaks will appear at higher frequencies than the original anticipated frequency. The leaks will then appear as an asymmetry at higher frequency with a decreasing amplitude when $l$ increases. This is indeed what is shown in Figure~\ref{Fig:simu}, where the observed ridge profiles obtained from PICARD/SODISM and SOHO/MDI intensity data are compared to the simulated profiles. From these simulations, we are able to understand:
\begin{itemize}
 \item why the peak in the $l^\prime$-$\nu$ diagram is shifted with $l$ (the shift with the original $l$-$\nu$  ridge increases as $l$ increases);
 \item why the peak gets closer to the original ridge as we get closer to the limb;
 \item why the  $l^\prime$-$\nu$ ridge leaks towards higher frequencies.
\end{itemize}

\begin{figure}[!t]
	\centerline{\includegraphics[width=0.5\textwidth]{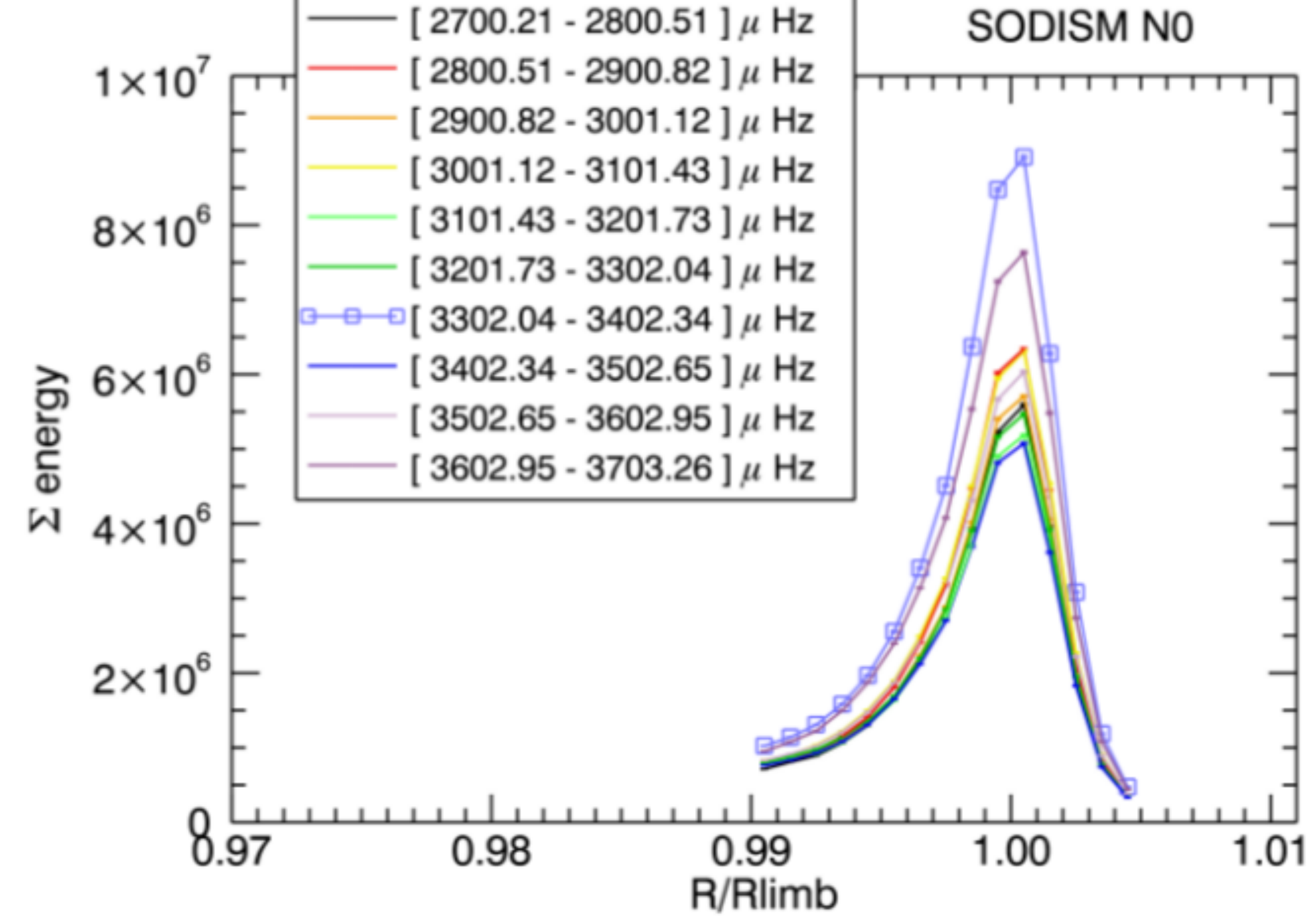}\includegraphics[width=0.5\textwidth]{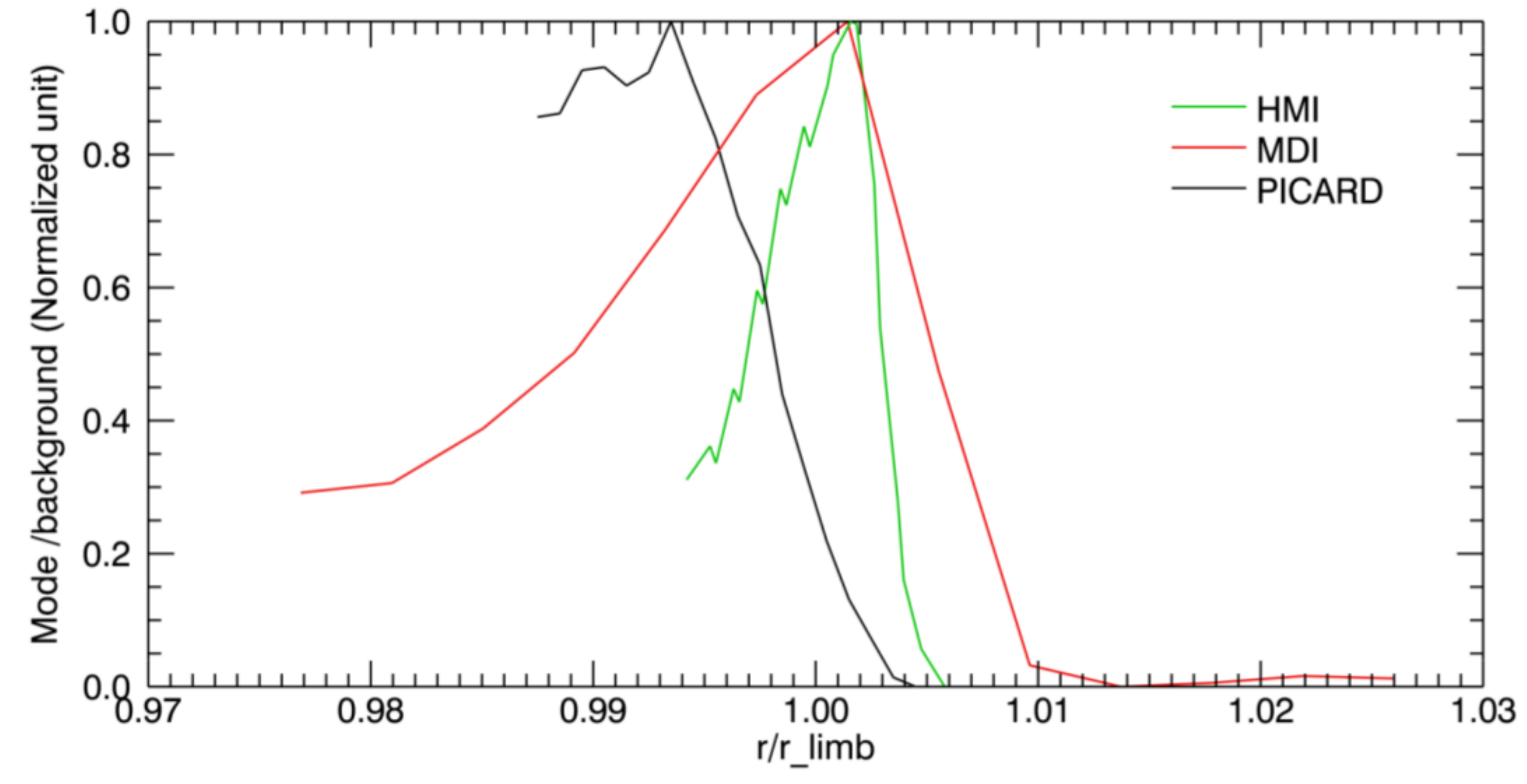}}
	\caption{Left: spectral energy level integrated in several frequency ranges of the 5-min oscillations envelope. Right: energy of the 5-min envelope normalized by the low frequency energy, for SODISM, SOHO/MDI and SDO/HMI. As far as SODISM is concerned, the amplification curve is displaced toward the inner part of the limb. This could be due to an instrumental effect not well taken into account in the data reduction pipeline. Note the narrower profile in the case of HMI, consequence of a better PSF.  }
	\label{Fig:amp}
\end{figure}

Figure~\ref{Fig:amp}  (left) shows   that the energy in the p-mode range peaks at the extreme limb, as was already seen using MDI intensity images \cite{Toner99} or the guiding pixels of SOHO/LOI \cite{App98b}. If we normalize this by a proxy of the noise level obtained by integrating the energy in the low-frequency band (right panel), we can see that we effectively have an amplification  of the signal-to-noise ratio at the extreme limb for the three instruments. This peak is however slightly shifted inward for SODISM. This may be an instrumental effect but this needs further investigations.

\section{Conclusions}
The first helioseismic analyses of SODISM full-continuum images have been carried out for both the medium-$l$ and limb programs. 
We  fitted mode parameters for a wide range of eigenmodes. The fitted peak asymmetries and frequency splittings are consistent with what is known from previous studies.
 We  performed a first inversion of the rotational splittings and  infered the internal rotation rate down to the radiative zone.
The  steep gradient of the tachocline is clearly seen at the base of the convection zone. 
 The signature of p-modes can also be observed in limb data and a signal amplification was detected at the extreme limb.  
By computing the response of any mode to the limb transform, we explain most of the features of the power spectra observed at the limb when compared to spectra built using full images. 
 However, we have to face strong instrumental and orbital effects that  affect both the photometric signal and the geometry of the solar images.   We know from the past that our capacity in assessing the reliability of
the details of many features analyzed using helioseismology 
(tachocline shape, polar jet, meridional flows, surface effects for structure inversions, etc.  see e.g. \cite{Schou02})
greatly benefit from the cross-analysis made using different instruments (e.g. GONG and SOHO/MDI).
 We are therefore still working on improving our calibration procedure and we hope that PICARD will be extended at least one year in order to gather more 
intensity data  that can be compared and cross-calibrated 
with SDO/HMI intensity  and velocity images.       

\ack
The data were provided by the PICARD/SODISM,  SOHO/MDI, and SDO/HMI instrument teams. PICARD is a mission supported by the Centre National d'Etudes Spatiales (CNES), the CNRS/INSU, the Belgian Space Policy (BELSPO), the Swiss Space Office (SSO) and the European Space Agency (ESA). SOHO is a project of international cooperation between ESA and NASA. D.~Salabert and P.~Journoud acknowledge financial support from CNES. The authors wish to thank S.~J.~Jim\'enez-Reyes for useful discussions and for providing his codes at the early stage of the project.

\end{document}